\documentclass[prl,aps,showpacs,twocolumn,amsmath,amssymb]{revtex4}

\usepackage{graphicx}
\usepackage{amssymb,amsmath}
\usepackage{color}

\begin{document}

\title{Reply to Comment on ``Quantum Entangled Dark Solitons formed by Ultracold Atoms in Optical Lattices''}
\author{R. V. Mishmash$^{1,2}$ and L. D. Carr$^{2}$}
\affiliation{$^1$Department of Physics, University of California, Santa Barbara, CA 93106, USA \\
$^2$Department of Physics, Colorado School of Mines, Golden, CO
80401, USA}
\date{\today}


\maketitle

In our entangled many-body quantum simulations we obtained three lines of evidence that mean-field-like dark solitons decay when taken as an initial condition: filling-in of the pair correlation function $g^{(2)}$; inelasticity of soliton-soliton collisions; and comparison of the exact quantum depletion of the macroscopic mode associated with the mean-field versus what one obtains with a weakly interacting Bogoliubov description.  The latter is covered in a great deal more detail in~\cite{carr2009m}.  Dziarmaga \textit{et al.} have showed that filling-in of $g^{(2)}$ does not necessarily correspond to filling-in or decay of a soliton in a single quantum measurement; that is, the soliton may delocalize, so that it fills in on average, but retain its soliton-like character in any single experiment, in keeping with well-established previous studies~\cite{dziarmaga2003}.  Martin and Ruostekoski have verified this result with an independent method~\cite{martinAD2010a, martinAD2010b}.

Dziarmaga \textit{et al.}'s simulation of a single-shot measurement treats the near-mean-field regime in which the filling factor for each site is $5000/31\simeq 161$.  They also work with an assumed mean-field soliton profile of a tanh function centered at different positions to build the delocalized soliton state.  Our work uses a filling factor of 1/2 to 2, and is not in the mean-field regime (although in the ground-state phase diagram our choices of hopping, interaction, and filling do correspond to the superfluid region, not the Mott-insulating one).  We agree with Dziarmaga \textit{et al.} that filling in of $g^{(2)}$ does not necessarily correspond to soliton decay in a single measurement.  However, the common difficulty of connecting results of near-mean-field simulations to entangled quantum many-body simulations remains; large filling factors have a different phenomenology than small filling factors, and one cannot use simulations from one regime to make a statement about simulations in the other.  Dziarmaga \textit{et al.}'s results are in a different regime from ours, and moreover have not dealt with our other two pieces of evidence for dark soliton decay.

In a separate work~\cite{carr2008a,carr2009f,carr2010c}, one of us has studied the quantum analog of dark solitons over all interaction regimes, from a mean field Bose-Einstein condensate to a Tonks-Girardeau gas. We showed that it is in fact yrast states that properly play the role of phase slip, not in general mean-field-like dark solitons.  These yrast states are precisely Lieb's type-II excitations, which were identified with dark solitons in the weakly-interacting limit only.  It is an intriguing open question as to whether or not anything like a dark soliton exists in the medium to strongly interacting regime, i.e., a well-defined, robust density notch which collides elastically with other solitons.  In either case, the mean-field dark soliton is very far from the quantum dark soliton, as we have shown in~\cite{carr2010c}, and thus we believe that mean-field dark solitons, e.g., Dziarmaga \textit{et al.}'s superposition of tanh condensates, decay when injected into the quantum theory out of the near-mean-field regime (we note that the initial conditions in our Letter are close to symmetry-broken versions of Dziarmaga \textit{et al.}'s tanh function).  The specific parameters for which mean-field theory breaks down for this problem were obtained via the Bethe ansatz, and the connection to the delocalization concept is discussed in more detail in~\cite{carr2010h}.

We thank Iacopo Carusotto, Rina Kanamoto, Carlos Lobo, Andrew D. Martin, Janne Ruostekoski, Alice Sinatra, and Masahito Ueda for useful discussions. We acknowledge support from the National Science Foundation under Grant PHY-0547845 as part of the NSF CAREER program.

\bibliographystyle{prsty}
%\bibliography{../../../refs/refs}
%\bibliography{refs}

\begin{thebibliography}{1}

\bibitem{carr2009m}
R.~V. Mishmash, I. Danshita, C.~W. Clark, and L.~D. Carr, Phys. Rev. A {\bf
  80},  053612  (2009).

\bibitem{dziarmaga2003}
J. Dziarmaga, Z.~P. Karkuszewski, and K. Sacha, J. Phys. B: At. Mol. Opt. Phys.
  {\bf 36},  1217  (2003).

\bibitem{martinAD2010a}
A.~D. Martin and J. Ruostekoski, Phys. Rev. Lett. {\bf 104}, 194102 (2010).

\bibitem{martinAD2010b}
A.~D. Martin and J. Ruostekoski, arXiv:1001.3385 (2010).

\bibitem{carr2008a}
R. Kanamoto, L.~D. Carr, and M. Ueda, Phys. Rev. Lett. {\bf 100},  060401
  (2008).

\bibitem{carr2009f}
R. Kanamoto, L.~D. Carr, and M. Ueda, Phys. Rev. A {\bf 79},  063616  (2009).

\bibitem{carr2010c}
R. Kanamoto, L.~D. Carr, and M. Ueda, Phys. Rev. A {\bf 81},  023625  (2010).

\bibitem{carr2010h}
L.~D. Carr, R. Kanamoto, and M. Ueda,  in {\em Understanding Quantum Phase
  Transitions}, edited by L.~D. Carr (Taylor \& Francis, Boca Raton, FL, 2010,
  to appear), Chap.~13.

\end{thebibliography}

\end{document}